\documentclass[twocolumn,letterpaper,aps,prd,superscriptaddress,longbibliography,floatfix]{revtex4-1}

\usepackage{graphicx}	
\usepackage{xspace}     

\begin{document}


\title{Transverse single spin asymmetries of forward neutrons in 
$p$$+$$p$, $p$$+$Al and $p$$+$Au collisions at $\sqrt{s_{_{NN}}}=200$ 
GeV as a function of transverse and longitudinal momenta}

\newcommand{\abilene}{Abilene Christian University, Abilene, Texas 79699, USA}
\newcommand{\augie}{Department of Physics, Augustana University, Sioux Falls, South Dakota 57197, USA}
\newcommand{\banaras}{Department of Physics, Banaras Hindu University, Varanasi 221005, India}
\newcommand{\barc}{Bhabha Atomic Research Centre, Bombay 400 085, India}
\newcommand{\baruch}{Baruch College, City University of New York, New York, New York, 10010 USA}
\newcommand{\bnlcoll}{Collider-Accelerator Department, Brookhaven National Laboratory, Upton, New York 11973-5000, USA}
\newcommand{\bnlphys}{Physics Department, Brookhaven National Laboratory, Upton, New York 11973-5000, USA}
\newcommand{\caucr}{University of California-Riverside, Riverside, California 92521, USA}
\newcommand{\charlesczech}{Charles University, Ovocn\'{y} trh 5, Praha 1, 116 36, Prague, Czech Republic}
\newcommand{\cns}{Center for Nuclear Study, Graduate School of Science, University of Tokyo, 7-3-1 Hongo, Bunkyo, Tokyo 113-0033, Japan}
\newcommand{\colorado}{University of Colorado, Boulder, Colorado 80309, USA}
\newcommand{\columbia}{Columbia University, New York, New York 10027 and Nevis Laboratories, Irvington, New York 10533, USA}
\newcommand{\czechtech}{Czech Technical University, Zikova 4, 166 36 Prague 6, Czech Republic}
\newcommand{\debrecen}{Debrecen University, H-4010 Debrecen, Egyetem t{\'e}r 1, Hungary}
\newcommand{\elte}{ELTE, E{\"o}tv{\"o}s Lor{\'a}nd University, H-1117 Budapest, P{\'a}zm{\'a}ny P.~s.~1/A, Hungary}
\newcommand{\eszterhazy}{Eszterh\'azy K\'aroly University, K\'aroly R\'obert Campus, H-3200 Gy\"ongy\"os, M\'atrai \'ut 36, Hungary}
\newcommand{\ewha}{Ewha Womans University, Seoul 120-750, Korea}
\newcommand{\famu}{Florida A\&M University, Tallahassee, FL 32307, USA}
\newcommand{\fsu}{Florida State University, Tallahassee, Florida 32306, USA}
\newcommand{\gsu}{Georgia State University, Atlanta, Georgia 30303, USA}
\newcommand{\hiroshima}{Hiroshima University, Kagamiyama, Higashi-Hiroshima 739-8526, Japan}
\newcommand{\howard}{Department of Physics and Astronomy, Howard University, Washington, DC 20059, USA}
\newcommand{\ihepprot}{IHEP Protvino, State Research Center of Russian Federation, Institute for High Energy Physics, Protvino, 142281, Russia}
\newcommand{\illuiuc}{University of Illinois at Urbana-Champaign, Urbana, Illinois 61801, USA}
\newcommand{\inrras}{Institute for Nuclear Research of the Russian Academy of Sciences, prospekt 60-letiya Oktyabrya 7a, Moscow 117312, Russia}
\newcommand{\instpasczech}{Institute of Physics, Academy of Sciences of the Czech Republic, Na Slovance 2, 182 21 Prague 8, Czech Republic}
\newcommand{\isu}{Iowa State University, Ames, Iowa 50011, USA}
\newcommand{\jaea}{Advanced Science Research Center, Japan Atomic Energy Agency, 2-4 Shirakata Shirane, Tokai-mura, Naka-gun, Ibaraki-ken 319-1195, Japan}
\newcommand{\jeonbuk}{Jeonbuk National University, Jeonju, 54896, Korea}
\newcommand{\kek}{KEK, High Energy Accelerator Research Organization, Tsukuba, Ibaraki 305-0801, Japan}
\newcommand{\korea}{Korea University, Seoul 02841, Korea}
\newcommand{\kurchatov}{National Research Center ``Kurchatov Institute", Moscow, 123098 Russia}
\newcommand{\kyoto}{Kyoto University, Kyoto 606-8502, Japan}
\newcommand{\lawllnl}{Lawrence Livermore National Laboratory, Livermore, California 94550, USA}
\newcommand{\losalamos}{Los Alamos National Laboratory, Los Alamos, New Mexico 87545, USA}
\newcommand{\lund}{Department of Physics, Lund University, Box 118, SE-221 00 Lund, Sweden}
\newcommand{\lyon}{IPNL, CNRS/IN2P3, Univ Lyon, Université Lyon 1, F-69622, Villeurbanne, France}
\newcommand{\maryland}{University of Maryland, College Park, Maryland 20742, USA}
\newcommand{\mass}{Department of Physics, University of Massachusetts, Amherst, Massachusetts 01003-9337, USA}
\newcommand{\michigan}{Department of Physics, University of Michigan, Ann Arbor, Michigan 48109-1040, USA}
\newcommand{\muhlenberg}{Muhlenberg College, Allentown, Pennsylvania 18104-5586, USA}
\newcommand{\nara}{Nara Women's University, Kita-uoya Nishi-machi Nara 630-8506, Japan}
\newcommand{\natmephi}{National Research Nuclear University, MEPhI, Moscow Engineering Physics Institute, Moscow, 115409, Russia}
\newcommand{\newmex}{University of New Mexico, Albuquerque, New Mexico 87131, USA}
\newcommand{\nmsu}{New Mexico State University, Las Cruces, New Mexico 88003, USA}
\newcommand{\northcg}{Physics and Astronomy Department, University of North Carolina at Greensboro, Greensboro, North Carolina 27412, USA}
\newcommand{\ohio}{Department of Physics and Astronomy, Ohio University, Athens, Ohio 45701, USA}
\newcommand{\ornl}{Oak Ridge National Laboratory, Oak Ridge, Tennessee 37831, USA}
\newcommand{\orsay}{IPN-Orsay, Univ.~Paris-Sud, CNRS/IN2P3, Universit\'e Paris-Saclay, BP1, F-91406, Orsay, France}
\newcommand{\peking}{Peking University, Beijing 100871, People's Republic of China}
\newcommand{\pnpi}{PNPI, Petersburg Nuclear Physics Institute, Gatchina, Leningrad region, 188300, Russia}
\newcommand{\pusan}{Pusan National University, Pusan 46241, Korea}
\newcommand{\riken}{RIKEN Nishina Center for Accelerator-Based Science, Wako, Saitama 351-0198, Japan}
\newcommand{\rikjrbrc}{RIKEN BNL Research Center, Brookhaven National Laboratory, Upton, New York 11973-5000, USA}
\newcommand{\rikkyo}{Physics Department, Rikkyo University, 3-34-1 Nishi-Ikebukuro, Toshima, Tokyo 171-8501, Japan}
\newcommand{\saispbstu}{Saint Petersburg State Polytechnic University, St.~Petersburg, 195251 Russia}
\newcommand{\seoulnat}{Department of Physics and Astronomy, Seoul National University, Seoul 151-742, Korea}
\newcommand{\stonybrkc}{Chemistry Department, Stony Brook University, SUNY, Stony Brook, New York 11794-3400, USA}
\newcommand{\stonycrkp}{Department of Physics and Astronomy, Stony Brook University, SUNY, Stony Brook, New York 11794-3800, USA}
\newcommand{\tenn}{University of Tennessee, Knoxville, Tennessee 37996, USA}
\newcommand{\texsu}{Texas Southern University, Houston, TX 77004, USA}
\newcommand{\titech}{Department of Physics, Tokyo Institute of Technology, Oh-okayama, Meguro, Tokyo 152-8551, Japan}
\newcommand{\tsukuba}{Tomonaga Center for the History of the Universe, University of Tsukuba, Tsukuba, Ibaraki 305, Japan}
\newcommand{\vandy}{Vanderbilt University, Nashville, Tennessee 37235, USA}
\newcommand{\weizmann}{Weizmann Institute, Rehovot 76100, Israel}
\newcommand{\wigner}{Institute for Particle and Nuclear Physics, Wigner Research Centre for Physics, Hungarian Academy of Sciences (Wigner RCP, RMKI) H-1525 Budapest 114, POBox 49, Budapest, Hungary}
\newcommand{\yonsei}{Yonsei University, IPAP, Seoul 120-749, Korea}
\newcommand{\zagreb}{Department of Physics, Faculty of Science, University of Zagreb, Bijeni\v{c}ka c.~32 HR-10002 Zagreb, Croatia}
\affiliation{\abilene}
\affiliation{\augie}
\affiliation{\banaras}
\affiliation{\barc}
\affiliation{\baruch}
\affiliation{\bnlcoll}
\affiliation{\bnlphys}
\affiliation{\caucr}
\affiliation{\charlesczech}
\affiliation{\cns}
\affiliation{\colorado}
\affiliation{\columbia}
\affiliation{\czechtech}
\affiliation{\debrecen}
\affiliation{\elte}
\affiliation{\eszterhazy}
\affiliation{\ewha}
\affiliation{\famu}
\affiliation{\fsu}
\affiliation{\gsu}
\affiliation{\hiroshima}
\affiliation{\howard}
\affiliation{\ihepprot}
\affiliation{\illuiuc}
\affiliation{\inrras}
\affiliation{\instpasczech}
\affiliation{\isu}
\affiliation{\jaea}
\affiliation{\jeonbuk}
\affiliation{\kek}
\affiliation{\korea}
\affiliation{\kurchatov}
\affiliation{\kyoto}
\affiliation{\lawllnl}
\affiliation{\losalamos}
\affiliation{\lund}
\affiliation{\lyon}
\affiliation{\maryland}
\affiliation{\mass}
\affiliation{\michigan}
\affiliation{\muhlenberg}
\affiliation{\nara}
\affiliation{\natmephi}
\affiliation{\newmex}
\affiliation{\nmsu}
\affiliation{\northcg}
\affiliation{\ohio}
\affiliation{\ornl}
\affiliation{\orsay}
\affiliation{\peking}
\affiliation{\pnpi}
\affiliation{\pusan}
\affiliation{\riken}
\affiliation{\rikjrbrc}
\affiliation{\rikkyo}
\affiliation{\saispbstu}
\affiliation{\seoulnat}
\affiliation{\stonybrkc}
\affiliation{\stonycrkp}
\affiliation{\tenn}
\affiliation{\texsu}
\affiliation{\titech}
\affiliation{\tsukuba}
\affiliation{\vandy}
\affiliation{\weizmann}
\affiliation{\wigner}
\affiliation{\yonsei}
\affiliation{\zagreb}
\author{U.A.~Acharya} \affiliation{\gsu} 
\author{C.~Aidala} \affiliation{\michigan} 
\author{Y.~Akiba} \email[PHENIX Spokesperson: ]{akiba@rcf.rhic.bnl.gov} \affiliation{\riken} \affiliation{\rikjrbrc} 
\author{M.~Alfred} \affiliation{\howard} 
\author{V.~Andrieux} \affiliation{\michigan} 
\author{N.~Apadula} \affiliation{\isu} 
\author{H.~Asano} \affiliation{\kyoto} \affiliation{\riken} 
\author{B.~Azmoun} \affiliation{\bnlphys} 
\author{V.~Babintsev} \affiliation{\ihepprot} 
\author{N.S.~Bandara} \affiliation{\mass} 
% ...
\collaboration{PHENIX Collaboration}  \noaffiliation

\date{\today}


\begin{abstract}


In 2015 the PHENIX collaboration at the Relativistic Heavy Ion Collider
recorded $p$$+$$p$, $p$$+$Al, and $p$$+$Au collision data at center of
mass energies of $\sqrt{s_{_{NN}}}=200$ GeV with the proton beam(s)
transversely polarized.  At very forward rapidities $\eta>6.8$ relative
to the polarized proton beam, neutrons were detected either inclusively
or in (anti)correlation with detector activity related to hard
collisions. The resulting single spin asymmetries, that were previously
reported, have now been extracted as a function of the transverse
momentum of the neutron as well as its longitudinal momentum fraction
$x_F$. The explicit kinematic dependence, combined with the correlation
information allows for a closer look at the interplay of different
mechanisms suggested to describe these asymmetries, such as hadronic
interactions or electromagnetic interactions in ultra-peripheral
collisions, UPC. Events that are correlated with a hard collision indeed
display a mostly negative asymmetry that increases in magnitude as a
function of transverse momentum with only little dependence on $x_F$. In
contrast, events that are not likely to have emerged from a hard
collision display positive asymmetries for the nuclear collisions with a
kinematic dependence that resembles that of a UPC based model. Because
the UPC interaction depends strongly on the charge of the nucleus, those
effects are very small for $p$$+$$p$ collisions, moderate for $p$$+$Al
collisions, and large for $p$$+$Au collisions.

\end{abstract}

\maketitle

\section{Introduction}

Very forward particle production in hadronic collisions has gained 
interest in the past decades due to its connection to other fields of 
science such as ultra-high-energy cosmic air showers. Such processes can 
be studied in controlled collisions to better calibrate their behavior. 
Given that at very forward regions the momentum transfer scales are 
often too small to describe the scattering process by perturbative 
quantum chromodynamics, nonperturbative descriptions of the strong 
interaction are needed to describe these processes. The simplest 
description is given by the exchange of a meson between the two beam 
hadrons. In Regge theory, very forward (but not necessarily very hard) 
processes were described by the exchange of the lightest hadrons between 
the colliding nucleons or nuclei, most dominantly by pions. Such 
calculations describe the measured forward neutron cross 
sections~\cite{Flauger:1976ju} reasonably well~\cite{Kopeliovich:2008da}.

The well known forward neutron single spin asymmetries that were 
discovered in $p$$+$$p$ 
collisions~\cite{Bazilevsky:2006vd,Togawa:2008cca,Adare:2013ekj}, 
though, could not be directly explained by one-pion exchange, OPE, 
alone. However, by adding the interference between pions and $a_1$ 
resonances such single hadron exchange models could also create 
asymmetries~\cite{Kopeliovich:2011bx}. This picture seemed to describe 
the measured forward neutron asymmetries in $p$$+$$p$ collisions 
reasonably well. The discovery of very different asymmetries in 
proton-nucleus collisions~\cite{Aidala:2017cnz} came initially as a big 
surprise because the sign and the magnitude of the asymmetries were 
quite different than the expectations. Very rapidly after these 
discoveries the impact of other contributions, most importantly 
electromagnetic interactions in ultra-peripheral collisions (UPC) was 
realized~\cite{Mitsuka:2017czj}. UPC describe the process where one 
nucleon or nucleus emits photons that then interact with the particle 
of the other beam. As it is an electromagnetic interaction, it can 
still take place when the impact parameters between the colliding 
nucleons or nuclei become too large for the strong interaction to 
contribute.

The interplay of the UPC interaction and hadronic interactions qualitatively 
describes the measured neutron asymmetries in proton-proton ($p$$+$$p$), 
proton-aluminum ($p$$+$Al) and proton-gold ($p$$+$Au) collisions, taking into 
account that the electromagnetic UPC interaction strongly depends on the 
charge of the colliding nucleus.  The contributions based on pion-$a_1$ 
interference contain a nearly linearly rising transverse momentum $p_T$ 
dependence (relative to the beam direction) while the dependence on the 
longitudinal momentum fraction $x_F$ (normalized by the beam momentum) is very 
weak. In contrast, the UPC related asymmetries depend strongly on the 
nucleon resonances that can be produced out of the transversely polarized 
proton and their subsequent decay kinematics. As such, the asymmetries are 
expected to change as a function of the neutron $p_T$ and $x_F$. For this 
purpose this paper extracts the single spin asymmetries of forward neutrons 
as a function of $p_T$ and $x_F$ in $p$$+$$p$, $p$$+$Al and $p$$+$Au collisions. In the 
previous paper~\cite{Acharya:2020opv}, only the transverse momentum 
dependence of the single spin asymmetries in $p$$+$$p$ collisions was extracted. 
In addition, in each collision system the neutron single spin asymmetries 
are extracted for inclusive neutrons, as well as in (anti)coincidence with 
hard-collision sensitive detector activity. These correlated results provide 
additional insights into the hadronic and UPC interactions as they either 
enhance or suppress either one of the contributions.

This paper is organized as follows. Initially the various data sets, 
necessary Monte Carlo simulations and the PHENIX detector systems relevant 
for this analysis are presented in section II. Then the event and neutron 
selection are explained in section III and the asymmetry extraction and 
unfolding of the raw asymmetries are presented in section IV. The results 
are discussed in section V before the paper is summarized.

\section{Data-sets and Monte Carlo simulations}

In 2015 the PHENIX experiment has taken data colliding transversely 
polarized proton beams with also transversely polarized proton beams, as 
well as Al and Au beams. The PHENIX detector is described in detail in 
reference~\cite{Adcox:2003zm}. Here only the detectors relevant for the 
following results are presented. Forward neutrons are detected with the Zero 
Degree Calorimeter (ZDC) which comprise three modules of Cu-W alloy 
absorbers that are layered with optical fibers. Each module corresponds to 
1.7 nuclear interaction lengths or 51 radiation lengths and has a projected 
transverse size of 10 cm by 10 cm with respect to the beam direction. Each 
of these layers is tilted 45 degrees upward to optimize the \v{C}erenkov 
light collection. The ZDCs are located approximately 18 m up and downstream 
of the PHENIX interaction point but for the presented analysis only the 
results of the ZDC at the proton-going beam direction are reported. In $p$$+$$p$ collisions also the same ZDC as in $p$$+$A collisions has been analyzed, calculating the asymmetries relative to the outgoing proton beam's spin orientation only.  Due to 
the different rigidities of protons and nuclei, the beam direction of the 
proton beam is tilted in proton-nucleus collisions and the ZDCs have been 
moved to still be located at the nominal proton-beam center position. The 
ZDCs cover pseudo-rapidities of $\eta>6.8$ and have an energy resolution of 
about 20\% for 100 GeV neutrons. The neutron hit positions are measured at 
the approximate maximum of the hadronic shower development with scintillator 
strip detectors between the first and second absorber modules. The strips of 
this shower-max detector (SMD) have a projected width of 15 mm horizontally 
and vertically. The hit positions in each direction are evaluated by 
obtaining the deposited energy weighted average of all strip positions. The 
SMD position resolutions are $\approx$1~cm in each direction. In addition, 
the SMDs are used in PHENIX for local polarimetry to confirm the beam spin 
orientation is rotated in the desired direction by making use of the nonzero 
neutron asymmetries.  In the proton-nucleus collisions, hodoscopes 
upstream of the ZDC were used to reject charged particles that potentially 
spray into the ZDCs from the magnet that merges and separates the two beams. 
These detectors were not installed in the $p$$+$$p$ collisions period in 
2015 and charged backgrounds were subtracted using their background 
contributions from the 2008 running period where these charge-veto detectors 
were available.

The correlation analysis used the beam-beam counters (BBCs), which
are located at 144 cm up and downstream of the nominal interaction point 
and cover pseudo-rapidities of 3.1 $ < |\eta| < $ 3.9. The BBCs comprise
of 64 quartz crystals, each being connected to photomultiplier tubes, 
and are sensitive to predominantly hard interactions. They provide the 
beam interaction vertex and timing information for most events. Both 
ZDCs and BBCs are used for event triggering and the BBCs are generally 
the basis for more sophisticated triggers that include other detector 
subsystems.

Various Monte Carlo (MC) generators were prepared to evaluate the 
uncertainties in the production of very forward neutrons. All of them 
were then piped through full {\sc geant}3~\cite{Brun:1994aa} simulations 
that include the ZDCs, the BBCs, the beam pipe, as well as the magnetic 
field of the dipole magnet that joins and separates the two beams.  The 
three MC generators {\sc dpmjet}~\cite{Roesler:2000he}, 
{\sc pythia}6.2~\cite{Sjostrand:2001yu} and 
{\sc pythia}8.2~\cite{Sjostrand:2014zea} generally describe a large 
variety of measurements reasonably well, but usually are more successful 
with hard interactions. As mentioned in the introduction, the very 
forward neutron production is very successfully described by one-pion 
exchange. Using an extracted two-dimensional very forward neutron 
distribution, neutrons are randomly created and the momentum and energy 
balance to the incoming proton is collided as a pion with the other beam 
using {\sc pythia}8.2 again. This allows one to create not only the very 
forward neutrons, but also the activity in the BBCs for the correlation 
studies. Last, the UPC interaction was simulated by creating the photon 
flux from the nuclear beam with the {\sc starlight}~\cite{Klein:2016yzr} 
generator and colliding it with the proton beam.

\section{Event and particle selection criteria}

For the inclusive and BBC-vetoed neutron measurements, events were selected 
via the ZDC triggers that require activity in at least one of the ZDCs and a 
deposited energy above 15 GeV. For the BBC-tagged neutron measurements 
events have been triggered via the BBCs, requiring at least one tube to fire 
per side. Neutrons are detected by the ZDCs if the reconstructed energy from 
all three modules is between 40 and 120 GeV. To remove electromagnetic 
particles at least 3\% of the total energy has to be accumulated in the 
second module. At least one hit in each of the horizontal and vertical SMD 
strips is required to determine the hit position. The radial hit position of 
the neutron then has to be between 0.5 and 4 cm to avoid acceptance effects 
as well as to avoid ambiguities in the azimuthal angles. Using the hit 
radius $r$ and the reconstructed energy $E$, the transverse momentum $p_T$ 
of the neutron can be extracted:
\begin{equation}
    p_{T} = \frac{r}{z_{ZDC} }E,
\end{equation}
where $z_{ZDC}$ is the $z$ position of the ZDC relative to the beam 
interaction point. The longitudinal momentum fraction is then calculated as:
\begin{equation}
    x_{F} = \frac{\sqrt{E^2-p_T^2}}{\sqrt{s_{_{NN}}}/2}\quad,
\end{equation}
in which the longitudinal momentum is normalized by the beam energy.
For the correlated measurements the BBC-tagged events require activity in at 
least one tube per side, while BBC-vetoed events were selected when neither 
side has any activity.

\section{Asymmetry extraction and unfolding}

The asymmetries are obtained by selecting neutron candidates in 4 transverse 
momentum bins from [0.01, 0.06, 0.11, 0.16, 0.21] GeV/$c$ plus one bin below 
([0.00, 0.01] GeV/$c$) and above ([0.21, 0.40] GeV/$c$) for the unfolding 
procedure. In $x_F$ also 4 bins from [0.4, 0.55, 0.7, 0.86, 1.0] GeV/$c$ are 
considered with again two boundary bins below ([0.0, 0.4] GeV/$c$) and above 
([1.0, 1.2] GeV/$c$) that only can be filled either by the true MC or the 
smeared data, respectively. The azimuthal angles are binned in 6 equidistant 
bins that cover the full azimuth where the angle is calculated with respect 
to the spin-up direction of the polarized proton beam. The asymmetries are 
then evaluated in each of these bins according to
\begin{equation}
    A_N(\phi) = \frac{1}{P} \frac{ N^\uparrow(\phi) - \mathcal{R} N^\downarrow(\phi)}
    { N^\uparrow(\phi) + \mathcal{R} N^\downarrow(\phi)}\quad, 
\end{equation}
where $N^{\uparrow},N^{\downarrow}$ are the neutron yields with the 
polarized proton beam spin orientation up or down, respectively. 
$\mathcal{R}$ represents the ratio of accumulated luminosities in these spin 
orientations that the PHENIX experiment sampled and is close to unity. $P$ 
is the average beam polarization which is 51.5\%, 59.4\%, and 59.1\% for $p$$+$$p$, 
$p$$+$Al, and $p$$+$Au collisions, respectively. In the case of $p$$+$$p$ collisions the spin orientation of the other beam was averaged over. The asymmetries are then fit with a 
sine modulation where for systematic studies phase and absolute 
normalization were left to vary as well.

The raw asymmetries were confirmed by using the so-called square-root method: 
\begin{equation}
    A_N = \frac{1}{P} \frac{\sqrt{N^\uparrow(\phi) N^\downarrow(\phi+\pi)} - \sqrt{N^\downarrow(\phi) N^\uparrow(\phi+\pi)}}{ \sqrt{N^\uparrow(\phi) N^\downarrow(\phi+\pi)} + \sqrt{N^\downarrow(\phi) N^\uparrow(\phi+\pi)}}\quad ,
\end{equation}
where the geometric mean of left/right yields in the detector cancels out 
differences in detector acceptance as well as differences in luminosities 
between the two spin orientations. Both results were found to be consistent 
with each other and no systematic uncertainty was assigned to it.

Due to the energy and position smearing, the kinematic values need to be 
unfolded. Unlike the previous publication~\cite{Acharya:2020opv}, for 
the inclusive neutron asymmetries in $p$$+$$p$ collisions, the unfolding 
was performed using the Bayes unfold method~\cite{DAgostini:1994fjx} as 
implemented in one (of various) methods in the 
RooUnfold~\cite{Adye:2011gm} package of Root~\cite{Brun:1997pa}. The 
advantage of using this unfolding package is that three-dimensional 
unfolding is directly available which is needed in this analysis. Also, 
the fact that the spin dependent yields (which implicitly contain the 
asymmetries) will not a priori be very similar to the (unpolarized) MC 
truth distributions is iteratively addressed in the Bayes unfolding. 
Therefore, one does not need to artificially create spin dependent true 
MC yields by re-weighting the MC according to various asymmetry 
assumptions. As a consequence, the large uncertainties related to 
reasonable asymmetry parameterizations used in the re-weighting for the 
TSVDunfolding~\cite{Hocker:1995kb} of the previous 
publication~\cite{Acharya:2020opv} are not needed here, and the result is 
substantially reduced overall uncertainties. In this paper the azimuthal 
angle $\phi$, the transverse momentum $p_T$ and the longitudinal 
momentum fraction $x_F$ are simultaneously unfolded on the 
spin-dependent yield level. From the unfolded spin dependent yields the 
resulting single spin asymmetries were again calculated as described 
above.

Following the Bayes-unfolding documentation on optimizing the unfolding, 
the number of iterations was varied around the value of four iterations 
by one. The variation in the resulting asymmetries was assigned as a
systematic uncertainty due to the unfolding procedure. Because the unfolding 
converges very rapidly, this uncertainty is nearly negligible in the 
total systematic uncertainties. Similar to the previous 
publication~\cite{Acharya:2020opv}, the variation of the unfolded 
asymmetries when using the different MC generators was also assigned as 
a systematic uncertainty. However, given that this unfolding method is 
much less sensitive to the MC input distributions these uncertainties 
are rather minor for most asymmetries. Only the impact of statistics at 
the edges of the dedicated MCs are visible. For the correlated 
asymmetries the unfolding was performed either on the full MC samples or 
applying the correlation selection also for the MC samples. This 
variation was also included as systematic uncertainty and reached up to 
a few percent.

Further sources of uncertainties were charged particle backgrounds in the 
$p$$+$$p$ collisions where the uncertainties on the relative background 
fractions were assigned to the spin-dependent yields. The variations in 
yields were also unfolded to result in uncertainties on the 
asymmetries. Next, the projected beam center position relative to the center 
of the ZDCs was varied by one cm in the horizontal and vertical directions. 
This variation was motivated by the position resolution of the SMD and  
potential changes in the beam conditions. The corresponding yields were then 
recalculated, unfolded and the variation of the resulting asymmetries' variation was again
assigned as a systematic uncertainty. In many kinematic bins, this is 
the dominating uncertainty of this measurement.  Lastly, as the asymmetries 
were normalized by the beam polarizations, the uncertainty on these values 
also enters the asymmetries. For the 2015 running period it was evaluated by 
the Relativistic-Heavy-Ion-Collider polarimetry group to be 3.4\%~\cite{polarimetry}.

The central values in $p_T$ and $x_F$ of the asymmetries do not contain any remaining uncertainties as the smearing was explicitly unfolded. Any differences potentially originating from the different MC generators are already absorbed in the corresponding systematic uncertainties described above. The central values in transverse momentum agree to the third digit between MC generators, while $x_F$ for the UPC generator differs up to 20\% from the other generators. 

Similar to the previous publication~\cite{Acharya:2020opv}, the OPE MC 
was artificially re-weighted to create single spin asymmetries. These 
weights were based on the following three functional forms: a third-order 
polynomial, a power-law behavior, and a negative exponential in the true 
neutron transverse momentum. These curves are not used in the 
unfolding, but are kept for comparison with the unfolded results. Again 
the curves that best reproduce the reconstructed asymmetries are 
expressed by lines for each functional form and the regions with a 
$\chi^2$ between reconstructed and smeared results below 40 are shaded 
as well. Because this re-weighting exercise was performed over the $x_F$ 
integrated data and MC, these curves also serve to compare the $x_F$ 
dependent variations.

\section{Results}

\begin{figure*}[htb]
\begin{minipage}{0.92\linewidth}
    \includegraphics[width=0.99\textwidth]{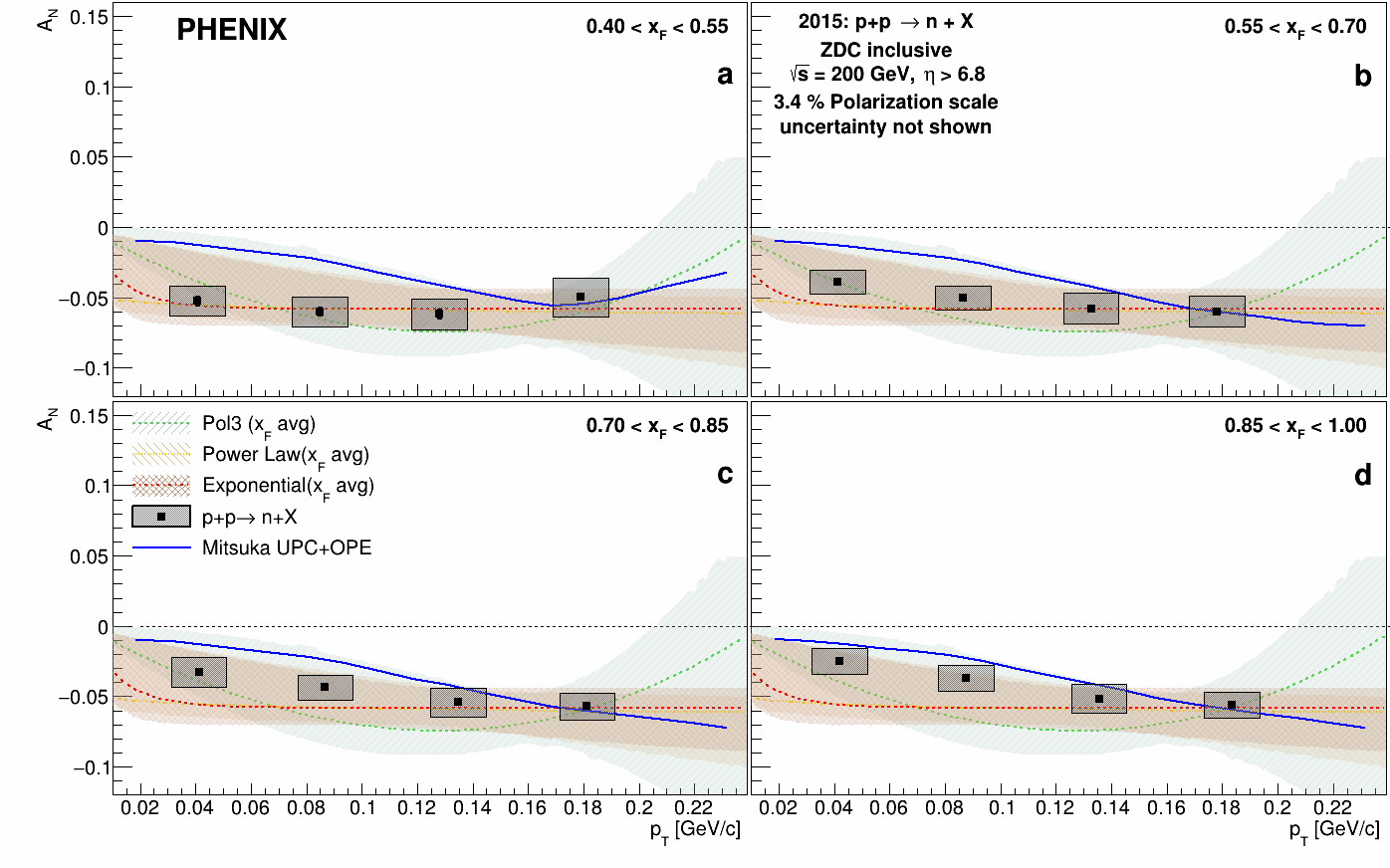}
\vspace{-0.5cm}
\caption{
Single spin asymmetries $A_N$ for very forward inclusive neutrons in $p$$+$$p$ 
collisions as a function of transverse momentum $p_T$, in bins of $x_F$. The 
error boxes represent the systematic uncertainties of the asymmetries while 
the solid [blue] lines represent the theory calculations. The size of uncertainty boxes does not reflect the bin boundaries and is chosen for visualization purposes. The three [colored] 
lines and shaded regions display the asymmetries from the reweighted MC that 
best describe the measured asymmetries integrated over all $x_F$ bins.}
    \label{fig:spec0_bbc0}
\end{minipage}
\begin{minipage}{0.92\linewidth}
\vspace{0.4cm}
    \includegraphics[width=0.99\textwidth]{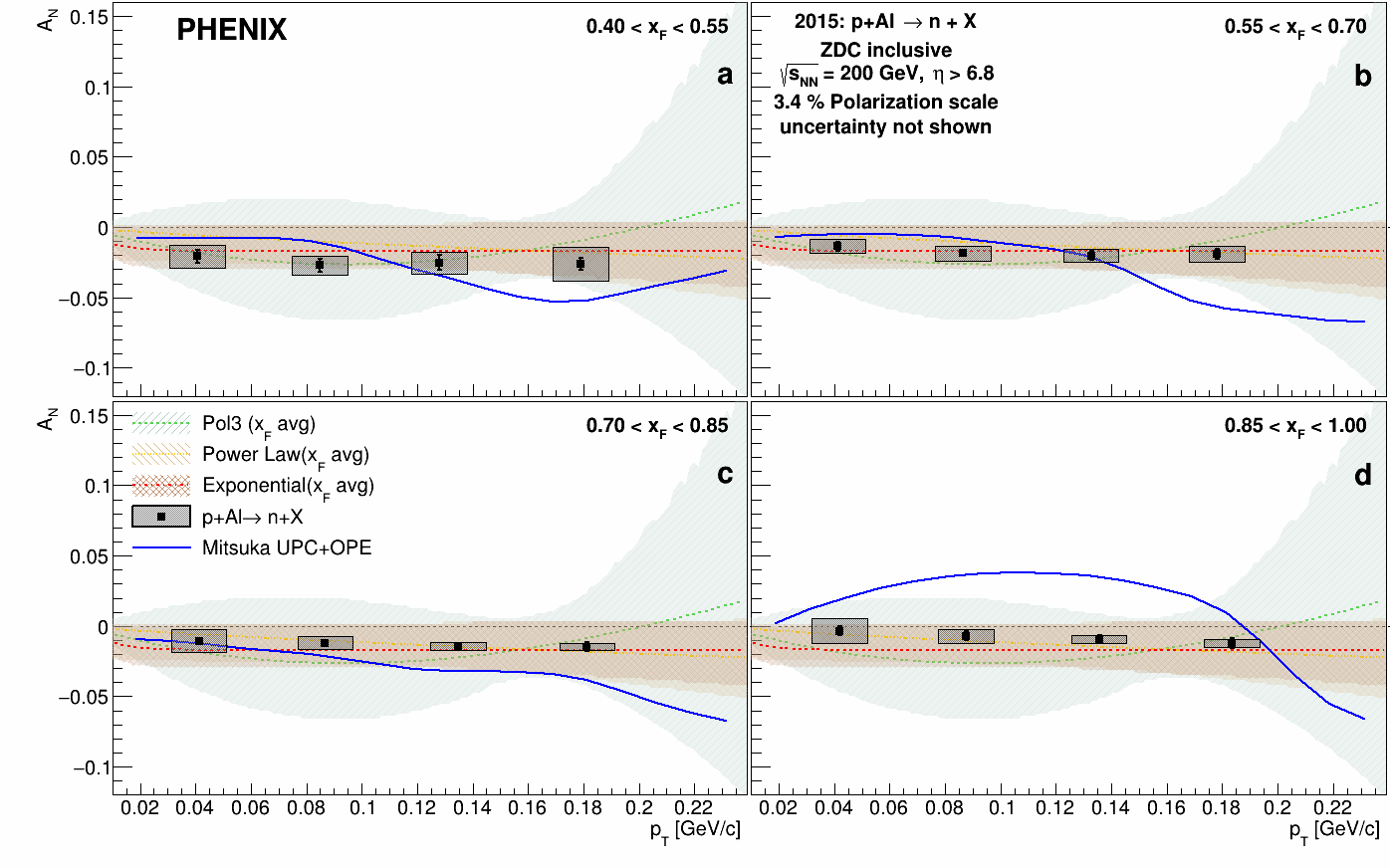}
\vspace{-0.5cm}
    \caption{
Single spin asymmetries $A_N$ for very forward inclusive neutrons in $p$$+$Al 
collisions as a function of transverse momentum $p_T$, in bins of $x_F$. The 
error boxes represent the systematic uncertainties of the asymmetries while 
the solid [blue] lines represent the theory calculations. The size of uncertainty boxes does not reflect the bin boundaries and is chosen for visualization purposes. The three [colored] 
lines and shaded regions display the asymmetries from the reweighted MC that 
best describe the measured asymmetries integrated over all $x_F$ bins.}
    \label{fig:spec1_bbc0}
\end{minipage}
\end{figure*}

\begin{figure*}[htb]
    \includegraphics[width=0.99\textwidth]{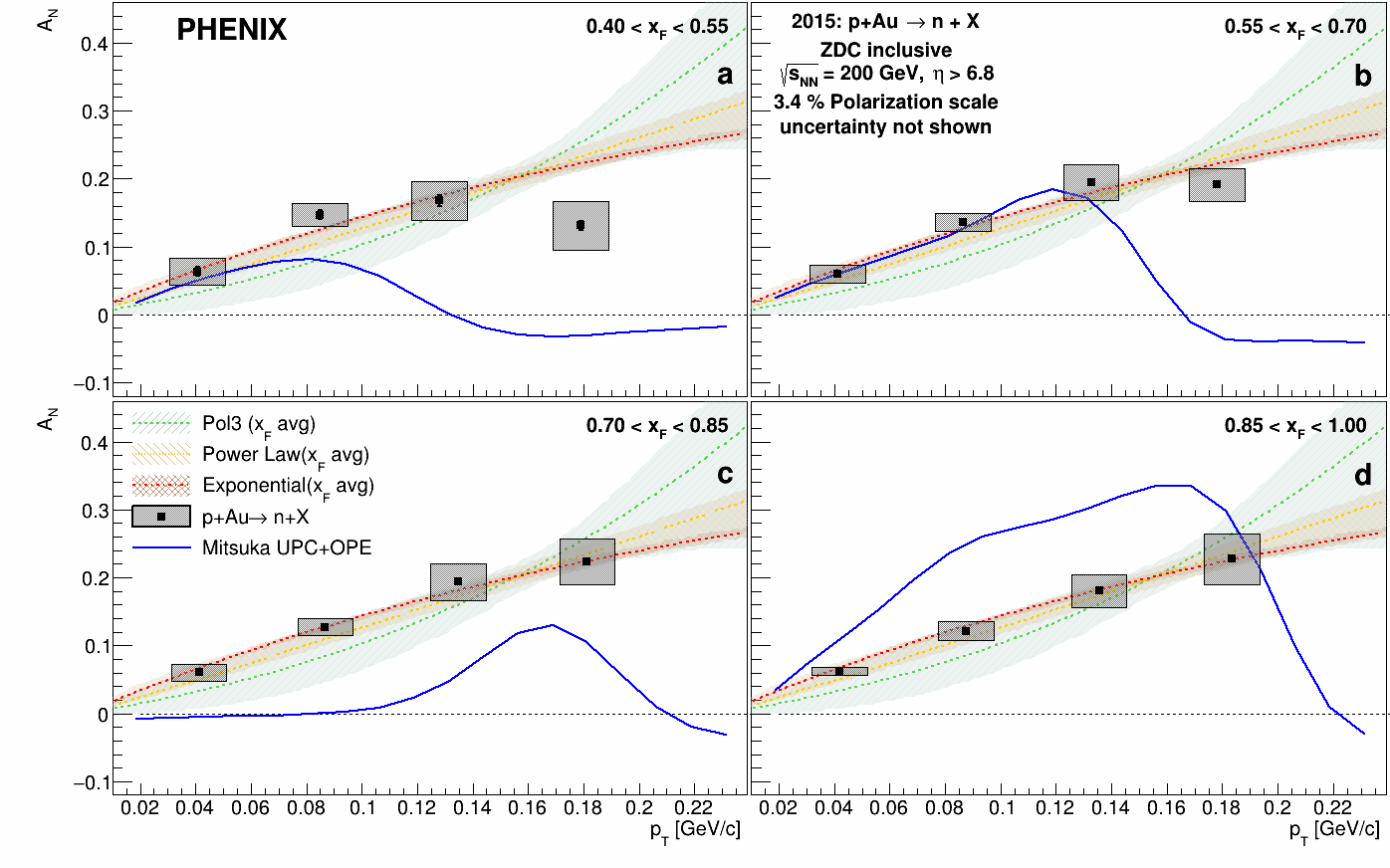}
\vspace{-0.5cm}
\caption{
Single spin asymmetries $A_N$ for very forward inclusive neutrons in $p$$+$Au 
collisions as a function of transverse momentum $p_T$, in bins of $x_F$. The 
error boxes represent the systematic uncertainties of the asymmetries while 
the solid [blue] lines represent the theory calculations. The size of uncertainty boxes does not reflect the bin boundaries and is chosen for visualization purposes. The three [colored] 
lines and shaded regions display the asymmetries from the reweighted MC that 
best describe the measured asymmetries integrated over all $x_F$ bins.}
    \label{fig:spec2_bbc0}
\end{figure*}

\begin{figure*}[htb]
\begin{minipage}{0.99\linewidth}
\vspace{-0.5cm}
    \includegraphics[width=0.9\textwidth]{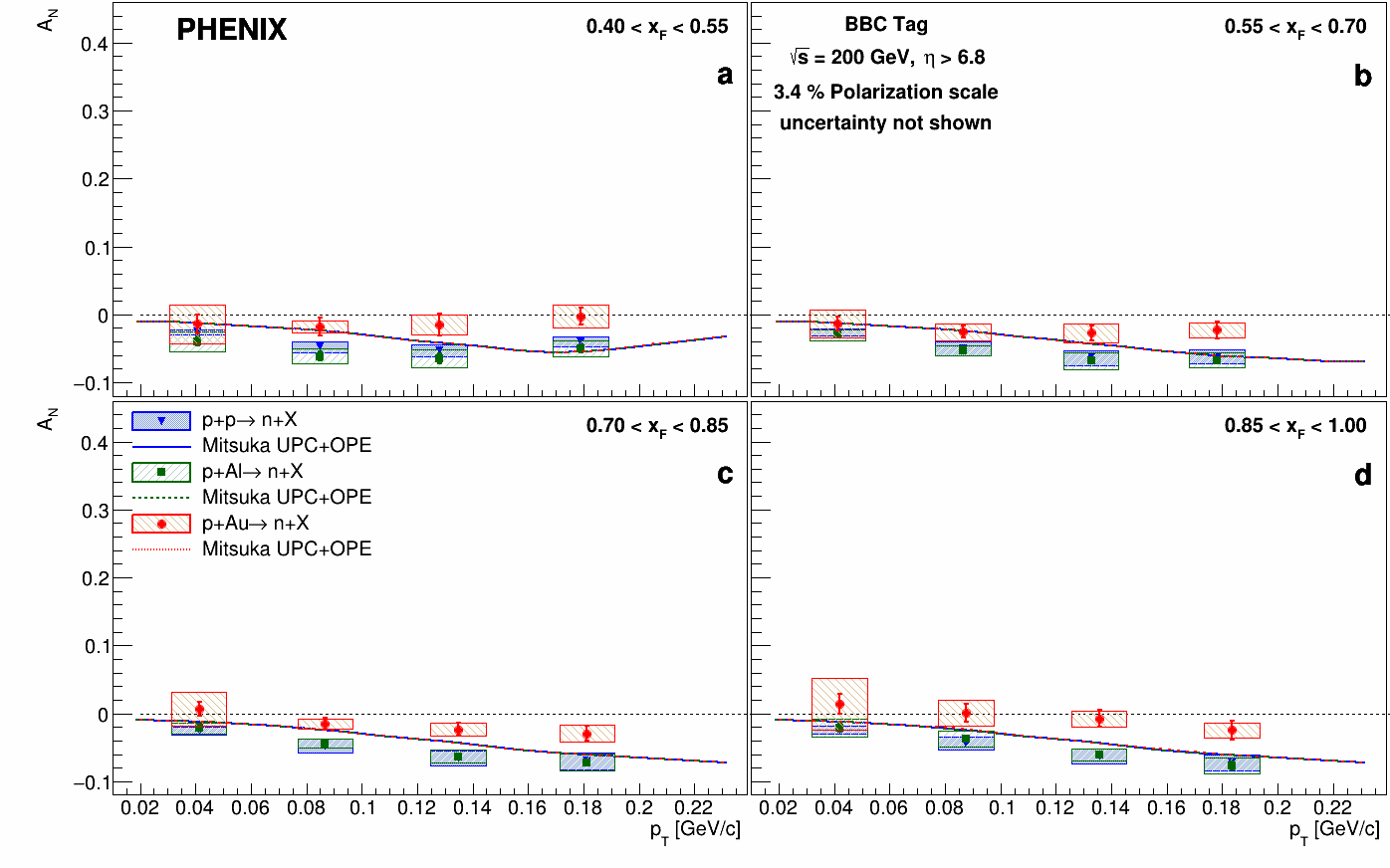}
\vspace{-0.5cm}
    \caption{
Single spin asymmetries $A_N$ for very forward neutrons in BBC tagged 
events in $p$$+$$p$ ([blue] triangles and boxes), $p$$+$Al ([green] squares and boxes) and 
$p$$+$Au ([red] circles and boxes) collisions as a function of transverse momentum 
$p_T$, in bins of $x_F$. The error boxes represent the systematic 
uncertainties of the asymmetries. The size of uncertainty boxes does not reflect the bin boundaries and is chosen for visualization purposes. The three [colored] lines represent the 
theory expectations for $p$$+$$p$ ([blue] lines), $p$$+$Al ([green] dashed 
lines) and $p$$+$Au ([red] dotted lines) as discussed in the text.}
    \label{fig:merged_bbc1}
\end{minipage}
\begin{minipage}{0.99\linewidth}
\vspace{0.5cm}
    \includegraphics[width=0.9\textwidth]{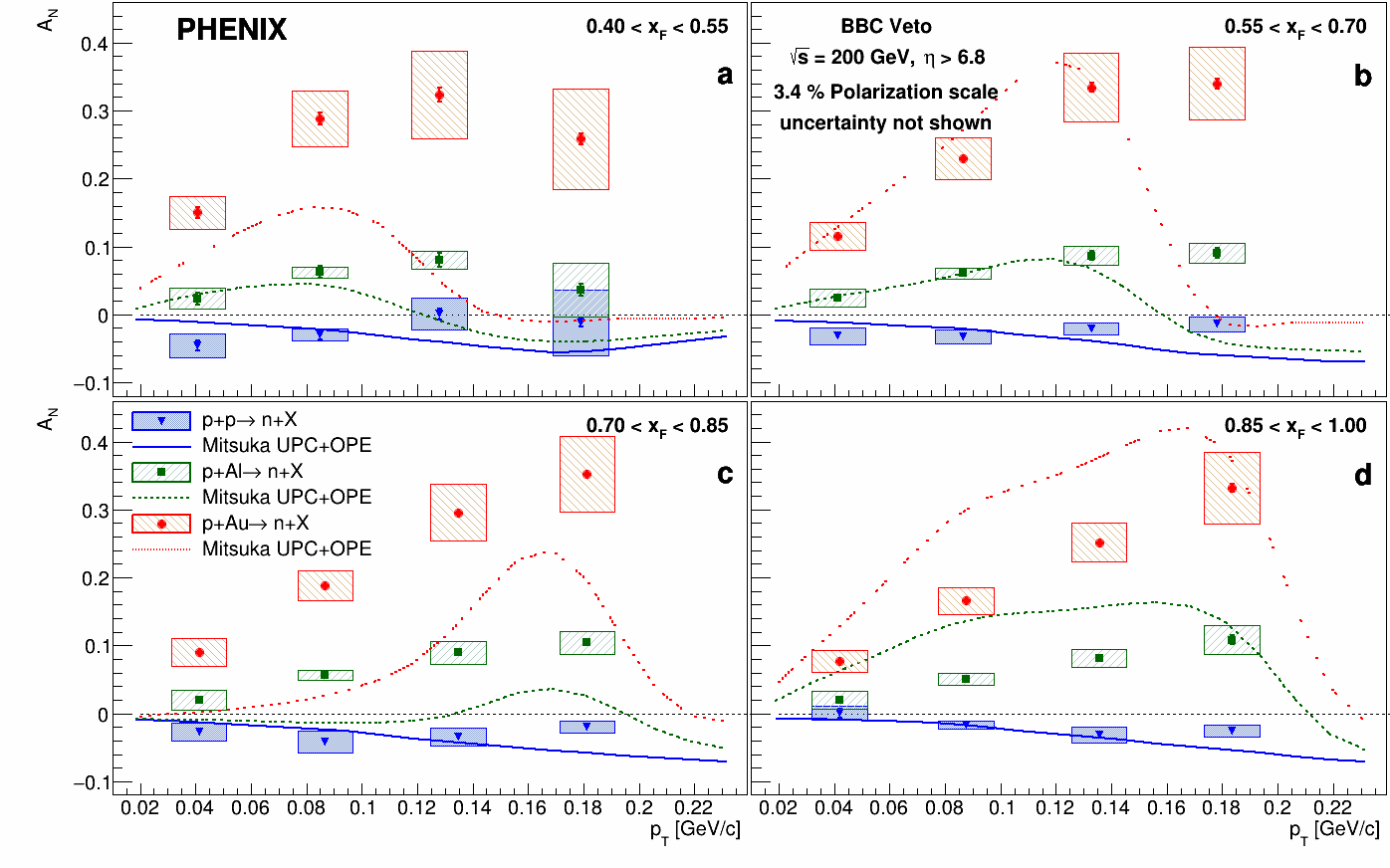}
\vspace{-0.5cm}
    \caption{
Single spin asymmetries $A_N$ for very forward neutrons in BBC-vetoed 
events in $p$$+$$p$ ([blue] triangles and boxes), $p$$+$Al ([green] 
squares and boxes) and $p$$+$Au ([red] circles and boxes) collisions as 
a function of transverse momentum $p_T$, in bins of $x_F$. The error 
boxes represent the systematic uncertainties of the asymmetries. The size of uncertainty boxes does not reflect the bin boundaries and is chosen for visualization purposes. The 
three [colored] lines represent the theory expectations for $p$$+$$p$ 
([blue] lines), $p$$+$Al ([green] dashed lines) and $p$$+$Au ([red] dotted 
lines) as discussed in the text.}
    \label{fig:merged_bbc2}
\end{minipage}
\end{figure*}

\begin{figure*}[htb]
\begin{minipage}{0.92\linewidth}
\vspace{-0.5cm}
    \includegraphics[width=0.95\textwidth]{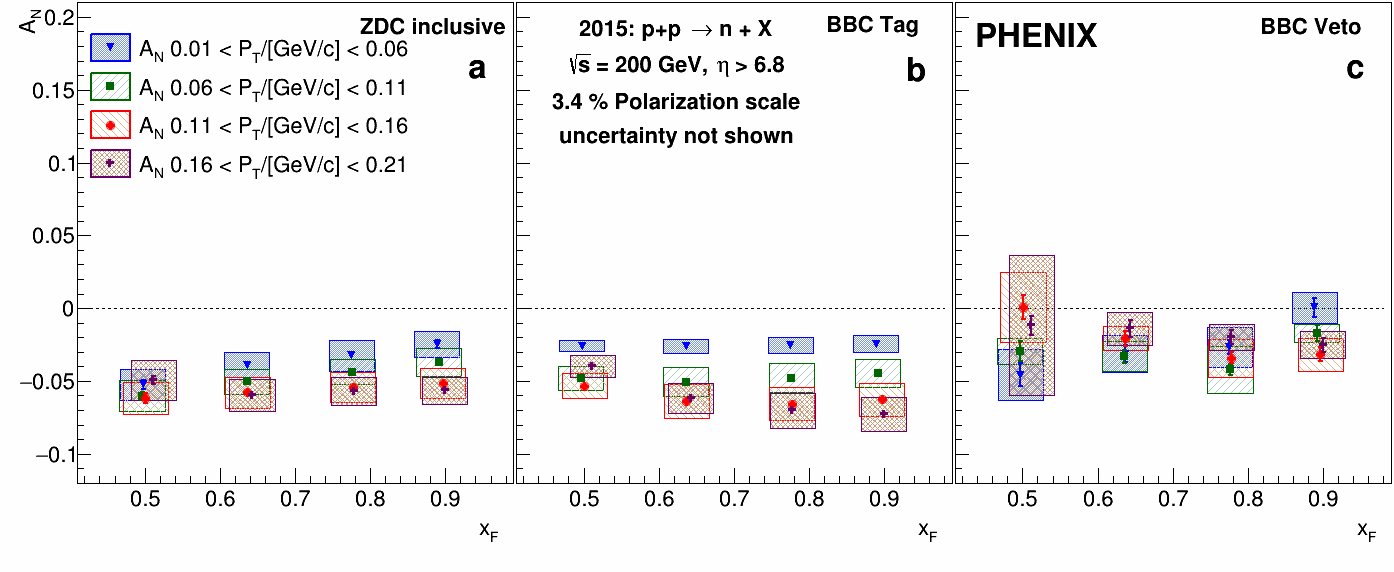}
\vspace{-0.5cm}
    \caption{
Single spin asymmetries $A_N$ for very forward neutrons in $p$$+$$p$ collisions as 
a function of $x_F$, in bins of transverse momentum $p_T$ ([blue] triangles, 
[green] squares, [red] circles, and [purple] crosses in ascending order). The 
error boxes represent the systematic uncertainties of the asymmetries. The size of uncertainty boxes does not reflect the bin boundaries and is chosen for visualization purposes. 
The three panels display (a) inclusive neutron asymmetries, (b) BBC-tagged 
neutron asymmetries, and (c) BBC-vetoed neutron asymmetries.}
    \label{fig:mergedxfpp}
\end{minipage}
\begin{minipage}{0.92\linewidth}
\vspace{0.5cm}
   \includegraphics[width=0.95\textwidth]{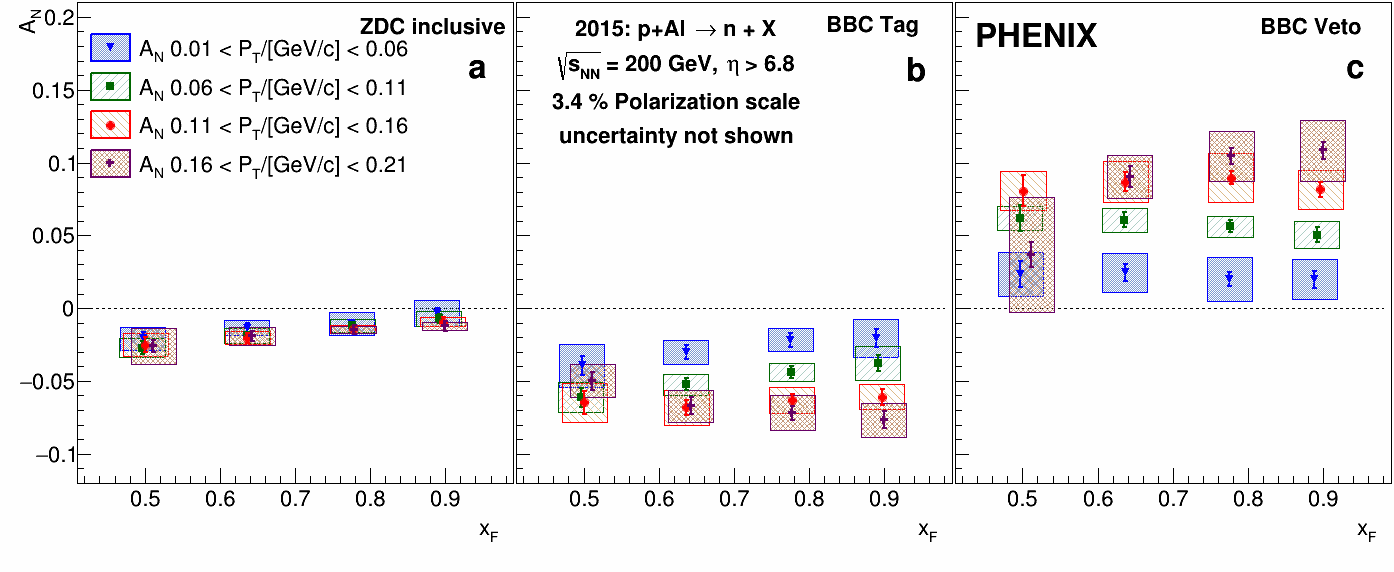}
\vspace{-0.5cm}
    \caption{
Single spin asymmetries $A_N$ for very forward neutrons in $p$$+$Al 
collisions as a function of $x_F$, in bins of transverse momentum $p_T$ 
([blue] triangles, [green] squares, [red] circles, and [purple] crosses in 
ascending order). The error boxes represent the systematic uncertainties 
of the asymmetries. The size of uncertainty boxes does not reflect the bin boundaries and is chosen for visualization purposes. The three panels display (a) inclusive neutron 
asymmetries, (b) BBC-tagged neutron asymmetries, and (c) BBC-vetoed 
neutron asymmetries.}
    \label{fig:mergedxfpAl}
\end{minipage}
\begin{minipage}{0.92\linewidth}
\vspace{0.5cm}
    \includegraphics[width=0.95\textwidth]{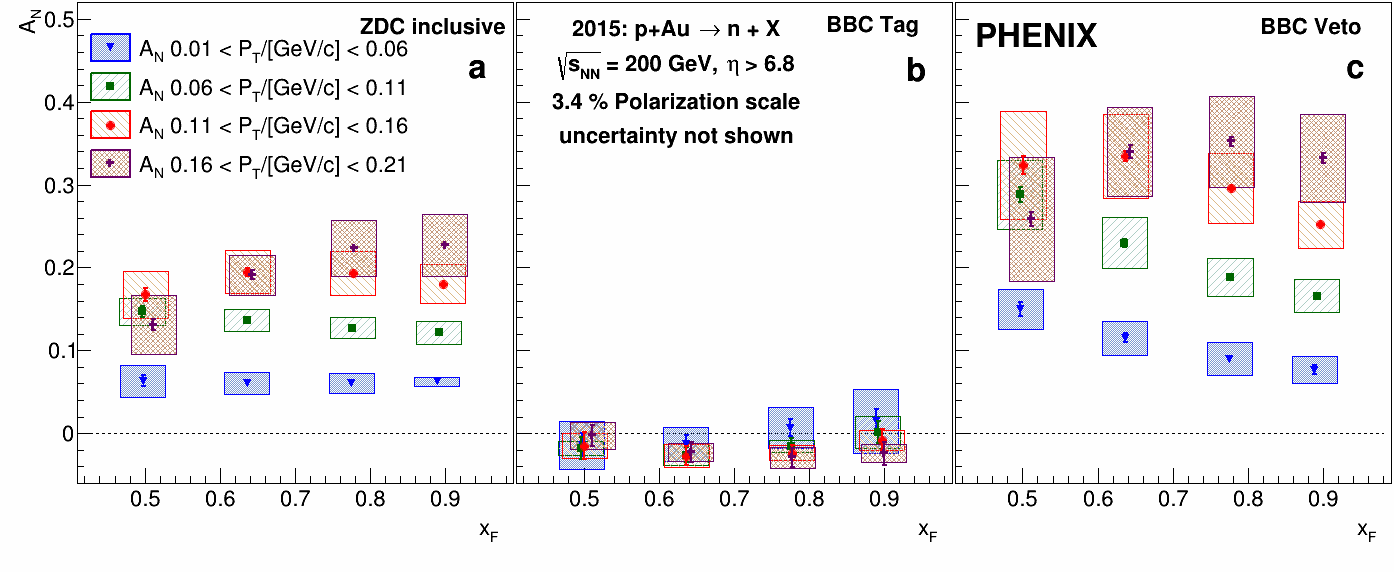}
\vspace{-0.5cm}
    \caption{
Single spin asymmetries $A_N$ for very forward neutrons in $p$$+$Au 
collisions as a function of $x_F$, in bins of transverse momentum $p_T$ 
([blue] triangles, [green] squares, [red] circles, and [purple] crosses 
in ascending order). The error boxes represent the systematic 
uncertainties of the asymmetries. The size of uncertainty boxes does not reflect the bin boundaries and is chosen for visualization purposes. The three panels display (a) inclusive 
neutron asymmetries, (b) BBC-tagged neutron asymmetries, and (c) 
BBC-vetoed neutron asymmetries.}
    \label{fig:mergedxfpAu}
\end{minipage}
\end{figure*}

Before discussing the final results, the theory calculations that are 
used for comparison of the asymmetries need to be discussed. The 
calculations comprise the aforementioned hadronic interactions of 
Ref.~\cite{Kopeliovich:2011bx} and the UPC based calculations of 
Ref.~\cite{Mitsuka:2017czj}. In Ref.~\cite{gakufine} both contributions have 
been combined, taking into account the mass number dependence of the 
hadronic interaction from Ref.~\cite{Guzey:2005tk} as described 
in Ref.~\cite{Mitsuka:2015twa}. In these calculations, the unpolarized 
neutron cross sections for OPE interactions are estimated to be 2.1 mb 
for $p$$+$$p$ collisions, 8.3 mb for $p$$+$Al collisions and 9.4 mb for 
$p$$+$Au collisions for neutron rapidities above 6.9 and $x_F>0.4$. In 
contrast, the estimated UPC-based contributions are below 0.1 mb for 
$p$$+$$p$ collisions, 0.7 mb for $p$$+$Al collisions and 19.6 mb for 
$p$$+$Au collisions~\cite{Mitsuka:2017czj}.

The calculated asymmetries for hadronic interactions are generally 
negative and increasing in magnitude with transverse momentum and 
essentially display no $x_F$ dependence. The UPC-based predictions 
display a more subtle structure based on the resonances that can be 
produced initially and their corresponding decay kinematics. At lower 
transverse momenta the $\Delta$ resonance dominantly contributes and 
creates positive asymmetries overall that are increasing with transverse 
momentum. At higher transverse momenta other resonances and their 
three-body decays may also contribute, but are not included in the 
polarized theory calculation. Therefore the theory calculations predict 
lower asymmetries. Also the $x_F$ dependence of the UPC-based asymmetry 
calculations appears to be varying to some extent.  

The current {\sc maid}-based asymmetry 
projections~\cite{Drechsel:2007if} are limited to invariant masses below 
2 GeV/$c^2$ and resonance decays into a single pion in addition to the 
neutron while other resonances and decays likely also contribute (as 
seen, for example, in the unpolarized {\sc starlight}-based 
simulations). As a consequence the sharp asymmetry behavior in the 
studied phase-space may be smoother and in particular may not exhibit 
such a sharp drop of the asymmetries at slightly higher transverse 
momenta. In the case of the correlated asymmetries, in addition to these 
cross section and asymmetry calculations also their relative 
reconstruction efficiencies need to be taken into account. These 
efficiencies are taken from the full {\sc geant}3 simulations. Within 
the OPE simulations, the efficiencies to fulfill the BBC tagging 
requirement are predominantly flat in neutron transverse momentum and 
drop from nearly 65\% at small $x_F$ to close to zero when $x_F$ 
approaches unity. Conversely, the BBC-veto efficiencies range from 
15\%--20\% at low $x_F$ to about 35\% close to unity. Overall, within 
this generator 35\% and 17\% of all detected neutron events get selected 
by BBC-tag and BBC-veto, respectively. Note that the remainder of events 
can be found having only one side of the BBCs showing activity. The 
other hadronic simulations also show a similar behavior with the actual 
numbers varying by several percent. In contrast, the UPC simulations 
have almost no activity in the BBCs with the BBC-veto efficiencies 
ranging around 85\%, and the BBC-tagging efficiencies are at or below 
5\% throughout the kinematic space covered in these measurements. 
Overall, 82\% of detected neutron events from this generator get 
selected in the BBC-vetoed case while the BBC-tagged fraction is below 
one percent. However, other backgrounds such as beam-gas 
interactions may still create random BBC activity.

Figure~\ref{fig:spec0_bbc0} shows the results for inclusive 
$p$$+$$p$ collisions as a function of transverse momentum and in bins of 
$x_F$. The asymmetries are negative in the whole kinematic region 
accessed by these measurements. They display a slightly increasing 
magnitude with increasing transverse momentum except for the lowest 
$x_F$ bin where the magnitude appears to decrease again. In $p$$+$$p$ 
collisions the interactions are dominated by the hadronic processes and 
hence the theory calculations are mostly negative and increasing in 
magnitude. Apart from the lowest $x_F$ bin, they describe the measured 
asymmetries well.

When studying the inclusive neutron asymmetries in $p$$+$Al collisions, 
the contribution from UPC events which follows the charge squared of the 
heavy nucleus will play a larger, but not yet dominating role in the 
interaction. As a consequence the mostly positive asymmetries from the 
UPC interaction and the mostly negative asymmetries from the hadron 
interaction cancel out to a large part as can be seen in 
Fig.~\ref{fig:spec1_bbc0}. Indeed, the asymmetries are only slightly 
negative and are often consistent with zero. The theory calculations are 
again describing the data mostly well.  At high $x_F$ some 
discrepancy is visible where a larger impact of the UPC events is 
predicted.

In the case of the $p$$+$Au collisions UPC events dominate the interaction and 
thus the asymmetries become positive as shown in Fig.~\ref{fig:spec2_bbc0}. 
The asymmetries are rising with increasing transverse momentum in all $x_F$ 
bins in a similar manner and reach more than 20\%. At lower $x_F$ there is 
an indication of the asymmetries saturating or even falling again which is 
qualitatively also seen in the theory predictions. The theory calculations 
show a substantial drop in the asymmetries at higher $p_T$ which is caused 
by moving away from the $\Delta$ resonance where the asymmetries decrease in 
the UPC calculations. The inclusion of other resonances in these 
calculations could contribute also in the asymmetries, as they have at least 
been shown to contribute in unpolarized calculations and the UPC MC 
generator. This likely also explains the smaller predicted asymmetries at 
lower $p_T$ in the $0.7<x_F<0.85$ region where $\Delta$ decays also cannot 
contribute. Again apart from these highest $p_T$ points, one can see that 
the $x_F$ dependence is mostly weak which can be seen when comparing with 
the $x_F$ averaged MC reweighting curves.

In the case of the correlated asymmetries the relative weights of the 
hadronic and ultra-peripheral interactions needed to be added to the theory 
calculations based on the OPE and UPC MC simulations mentioned above. 
Generally, a large fraction of hadronic events also fire the BBCs as there 
is activity in a wide range of rapidities in addition to the very forward 
region where the neutrons are detected. In contrast UPC interactions, where 
photons at relatively small scales only excite the proton to a $\Delta$ 
resonance or other nucleon states, very little activity other than from the 
decay products of these states is expected. Therefore, the overwhelming 
majority of UPC events do not show any activity in the BBCs. Consequently, 
the BBC-tagged events are dominated by hadronic interactions and the 
BBC-vetoed events have a larger UPC contribution depending on the relative 
contributions of the colliding systems.

Figure~\ref{fig:merged_bbc1} displays the BBC-tagged asymmetries as a 
function of transverse momentum in bins of $x_F$ for $p$$+$$p$, 
$p$$+$Al, and $p$$+$Au collisions, respectively. The $p$$+$$p$ and the 
$p$$+$Al asymmetries are indeed negative and show a very similar 
behavior. The asymmetries are increasing in magnitude more than the 
inclusive neutron asymmetries and reach magnitudes close to 10\%. A 
nearly linear behavior as predicted by~\cite{Kopeliovich:2011bx} appears 
to describe these asymmetries rather well as can be seen also in the 
combined theory calculations that match the data very well. In the 
lowest $x_F$ bin there is an indication that at higher $p_T$ the 
asymmetries become smaller again which actually is well described by the 
combined theory calculations and thus indicates some residual 
contribution from the UPC interaction. In $p$$+$Au collisions the 
asymmetries are generally smaller and only are slightly negative at 
intermediate $x_F$. This indicates that some UPC events do remain in 
$p$$+$Au collisions and dilute the predominantly hadronic asymmetries.

In the case of the BBC-vetoed asymmetries the contributions from UPC 
interactions are enhanced while those from hadronic interactions are reduced 
but not eliminated. In $p$$+$$p$ collisions these two contributions mostly cancel 
each other out as can be seen in Fig.~\ref{fig:merged_bbc2} where the 
asymmetries are close to zero in most kinematic regions. A hint of a 
negative asymmetry at lower transverse momenta is still visible but the 
asymmetries become smaller as the transverse momentum increases. In $p$$+$Al 
collisions, the larger UPC contribution now switches the sign of the 
asymmetries to become positive and rising in transverse momentum, reaching 
up to 10\% at the higher transverse momenta. The $p_T$ dependence is again 
mostly similar for all $x_F$ bins, but at lower $x_F$ the asymmetry in the 
highest transverse momentum appears to be dropping similar to the inclusive 
$p$$+$Au collision case. The theory calculations are again qualitatively 
comparable with the data, including this drop at higher transverse momenta.

In the BBC-vetoed neutron asymmetries of $p$$+$Au collisions the UPC 
contributions are even further dominant and the resulting asymmetries are 
even less diluted by the hadronic interactions. This can be seen by the 
asymmetries now even exceeding 30\%. Given the impact of the produced 
nucleon resonances on the decay kinematics, here the largest dependence on 
$x_F$ can be seen with slightly higher asymmetries at low $x_F$ than in 
higher bins. The drop of the asymmetries at high $p_T$ at lower $x_F$ noted 
already in the inclusive $p$$+$Au and the BBC-vetoed $p$$+$Al asymmetries is also 
visible here.

To look more closely at the $x_F$ dependence, Figs.~\ref{fig:mergedxfpp} to 
\ref{fig:mergedxfpAu} display the same asymmetries as a function of $x_F$ in 
bins of transverse momentum. In the $p$$+$$p$ collisions, within uncertainties, 
very little dependence on the longitudinal momentum fraction can be seen 
overall. The only trends can be seen at low transverse momenta for either 
inclusive or BBC-vetoed neutron events where the magnitude of the 
asymmetries becomes smaller with increasing $x_F$. For the BBC-tagged $p$$+$$p$ 
asymmetries, the magnitude for the lowest $x_F$ bin is smaller at higher 
transverse momenta.

For neutrons in $p$$+$Al collisions, the inclusive neutron asymmetries appear to 
be getting smaller in magnitude as a function $x_F$ for all transverse 
momentum bins. This could be an indication that the relative contribution of 
UPC events becomes stronger with $x_F$. A similar behavior is also visible 
for BBC-tagged neutrons, at least for smaller transverse momenta while it is 
mostly flat in $x_F$ at higher transverse momenta.

In $p$$+$Au collisions, the neutron asymmetries show again generally a weak 
dependence on $x_F$. Only in the BBC-vetoed events a clear trend can be seen 
for lower transverse momenta where the asymmetries decrease with increasing 
$x_F$. At higher transverse momenta the uncertainties become too large to 
tell whether this trend continues. As mentioned previously, because the UPC 
based asymmetries closely rely on the resonances that can be formed, this is 
where one can expect the largest variation of the asymmetries with either 
kinematic variable.

\section{Summary}

In summary, the PHENIX experiment has measured single transverse spin 
asymmetries of very forward neutrons as a function of transverse momentum 
and longitudinal momentum fraction $x_F$. The asymmetries were extracted in 
$p$$+$$p$, $p$$+$Al, and $p$$+$Au collisions and for neutrons that were either inclusively 
extracted or in (anti)correlation with hard collision-sensitive detectors. 
The asymmetries show a strong dependence on the collision system which can 
qualitatively be described by the interplay of hard interactions and 
ultra-peripheral collisions which strongly depend on the charge of the 
colliding nucleus. These indications were corroborated by the 
(anti)correlated results that either enhance or reduce the contributions of 
the two competing processes. The asymmetries in hadronic processes appear to 
be negative and increase in magnitude with increasing transverse momentum to 
up to 10\% while hardly any $x_F$ dependence is visible. The UPC related 
asymmetries in contrast are positive and reach more than 30\% in magnitude. 
Those are also initially rising with transverse momentum while for low $x_F$ 
a subsequent decrease of the asymmetries is seen. Also at lower transverse 
momenta some differences in $x_F$ are visible that in the model calculations 
originate from the different nucleon resonances contributing. The precision 
and the multi-dimensional dependence of these measurements will greatly 
improve the general understanding of how hadronic and photon-induced 
processes create single spin asymmetries at very forward rapidities and at 
small observed transverse-momentum scales.


\begin{acknowledgments} 

We thank the staff of the Collider-Accelerator and Physics
Departments at Brookhaven National Laboratory and the staff of
the other PHENIX participating institutions for their vital
contributions.  
We acknowledge support from the Office of Nuclear Physics in the
Office of Science of the Department of Energy,
the National Science Foundation,
Abilene Christian University Research Council,
Research Foundation of SUNY, and
Dean of the College of Arts and Sciences, Vanderbilt University
(U.S.A),
Ministry of Education, Culture, Sports, Science, and Technology
and the Japan Society for the Promotion of Science (Japan),
Natural Science Foundation of China (People's Republic of China),
Croatian Science Foundation and
Ministry of Science and Education (Croatia),
Ministry of Education, Youth and Sports (Czech Republic),
Centre National de la Recherche Scientifique, Commissariat
{\`a} l'{\'E}nergie Atomique, and Institut National de Physique
Nucl{\'e}aire et de Physique des Particules (France),
J. Bolyai Research Scholarship, EFOP, the New National Excellence
Program ({\'U}NKP), NKFIH, and OTKA (Hungary),
Department of Atomic Energy and Department of Science and Technology
(India),
Israel Science Foundation (Israel),
Basic Science Research and SRC(CENuM) Programs through NRF
funded by the Ministry of Education and the Ministry of
Science and ICT (Korea).
Ministry of Education and Science, Russian Academy of Sciences,
Federal Agency of Atomic Energy (Russia),
VR and Wallenberg Foundation (Sweden),
the U.S. Civilian Research and Development Foundation for the
Independent States of the Former Soviet Union,
the Hungarian American Enterprise Scholarship Fund,
the US-Hungarian Fulbright Foundation,
and the US-Israel Binational Science Foundation.

\end{acknowledgments}  




\begin{thebibliography}{24}%
\makeatletter
\providecommand \@ifxundefined [1]{%
 \@ifx{#1\undefined}
}%
\providecommand \@ifnum [1]{%
 \ifnum #1\expandafter \@firstoftwo
 \else \expandafter \@secondoftwo
 \fi
}%
\providecommand \@ifx [1]{%
 \ifx #1\expandafter \@firstoftwo
 \else \expandafter \@secondoftwo
 \fi
}%
\providecommand \natexlab [1]{#1}%
\providecommand \enquote  [1]{``#1''}%
\providecommand \bibnamefont  [1]{#1}%
\providecommand \bibfnamefont [1]{#1}%
\providecommand \citenamefont [1]{#1}%
\providecommand \href@noop [0]{\@secondoftwo}%
\providecommand \href [0]{\begingroup \@sanitize@url \@href}%
\providecommand \@href[1]{\@@startlink{#1}\@@href}%
\providecommand \@@href[1]{\endgroup#1\@@endlink}%
\providecommand \@sanitize@url [0]{\catcode `\\12\catcode `\$12\catcode
  `\&12\catcode `\#12\catcode `\^12\catcode `\_12\catcode `\%12\relax}%
\providecommand \@@startlink[1]{}%
\providecommand \@@endlink[0]{}%
\providecommand \url  [0]{\begingroup\@sanitize@url \@url }%
\providecommand \@url [1]{\endgroup\@href {#1}{\urlprefix }}%
\providecommand \urlprefix  [0]{URL }%
\providecommand \Eprint [0]{\href }%
\providecommand \doibase [0]{http://dx.doi.org/}%
\providecommand \selectlanguage [0]{\@gobble}%
\providecommand \bibinfo  [0]{\@secondoftwo}%
\providecommand \bibfield  [0]{\@secondoftwo}%
\providecommand \translation [1]{[#1]}%
\providecommand \BibitemOpen [0]{}%
\providecommand \bibitemStop [0]{}%
\providecommand \bibitemNoStop [0]{.\EOS\space}%
\providecommand \EOS [0]{\spacefactor3000\relax}%
\providecommand \BibitemShut  [1]{\csname bibitem#1\endcsname}%
\let\auto@bib@innerbib\@empty
\bibitem [{\citenamefont {Flauger}\ and\ \citenamefont
  {Monnig}(1976)}]{Flauger:1976ju}%
  \BibitemOpen
  \bibfield  {author} {\bibinfo {author} {\bibfnamefont {W.}~\bibnamefont
  {Flauger}}\ and\ \bibinfo {author} {\bibfnamefont {F.}~\bibnamefont
  {Monnig}},\ }\bibfield  {title} {\enquote {\bibinfo {title} {{Measurement of
  Inclusive Zero-Angle Neutron Spectra at the ISR}},}\ }\href {\doibase
  10.1016/0550-3213(76)90211-X} {\bibfield  {journal} {\bibinfo  {journal}
  {Nucl. Phys. B}\ }\textbf {\bibinfo {volume} {109}},\ \bibinfo {pages} {347}
  (\bibinfo {year} {1976})}\BibitemShut {NoStop}%
\bibitem [{\citenamefont {Kopeliovich}\ \emph {et~al.}(2008)\citenamefont
  {Kopeliovich}, \citenamefont {Potashnikova}, \citenamefont {Schmidt},\ and\
  \citenamefont {Soffer}}]{Kopeliovich:2008da}%
  \BibitemOpen
  \bibfield  {author} {\bibinfo {author} {\bibfnamefont {B.~Z.}\ \bibnamefont
  {Kopeliovich}}, \bibinfo {author} {\bibfnamefont {I.~K.}\ \bibnamefont
  {Potashnikova}}, \bibinfo {author} {\bibfnamefont {I.}~\bibnamefont
  {Schmidt}}, \ and\ \bibinfo {author} {\bibfnamefont {J.}~\bibnamefont
  {Soffer}},\ }\bibfield  {title} {\enquote {\bibinfo {title} {{Damping of
  forward neutrons in $p p$ collisions}},}\ }\href {\doibase
  10.1103/PhysRevD.78.014031} {\bibfield  {journal} {\bibinfo  {journal} {Phys.
  Rev. D}\ }\textbf {\bibinfo {volume} {78}},\ \bibinfo {pages} {014031}
  (\bibinfo {year} {2008})}\BibitemShut {NoStop}%
\bibitem [{\citenamefont {Fukao}\ \emph {et~al.}(2007)\citenamefont {Fukao}
  \emph {et~al.}}]{Bazilevsky:2006vd}%
  \BibitemOpen
  \bibfield  {author} {\bibinfo {author} {\bibfnamefont {Y.}~\bibnamefont
  {Fukao}} \emph {et~al.},\ }\bibfield  {title} {\enquote {\bibinfo {title}
  {{Single Transverse-Spin Asymmetry in Very Forward and Very Backward Neutral
  Particle Production for Polarized Proton Collisions at $\sqrt{s}=200$
  GeV}},}\ }\href {\doibase 10.1016/j.physletb.2007.05.049} {\bibfield
  {journal} {\bibinfo  {journal} {Phys. Lett. B}\ }\textbf {\bibinfo {volume}
  {650}},\ \bibinfo {pages} {325} (\bibinfo {year} {2007})}\BibitemShut
  {NoStop}%
\bibitem [{\citenamefont {Togawa}(2008)}]{Togawa:2008cca}%
  \BibitemOpen
  \bibfield  {author} {\bibinfo {author} {\bibfnamefont {M.}~\bibnamefont
  {Togawa}},\ }\emph {\bibinfo {title} {{Measurements of the leading neutron
  production in polarized $pp$ collision at $\sqrt{s}$=200 GeV}}},\ \href@noop
  {} {Ph.D. thesis},\ \bibinfo  {school} {Kyoto U.} (\bibinfo {year}
  {2008})\BibitemShut {NoStop}%
\bibitem [{\citenamefont {Adare}\ \emph {et~al.}(2014)\citenamefont {Adare}
  \emph {et~al.}}]{Adare:2013ekj}%
  \BibitemOpen
  \bibfield  {author} {\bibinfo {author} {\bibfnamefont {A.}~\bibnamefont
  {Adare}} \emph {et~al.} (\bibinfo {collaboration} {PHENIX Collaboration}),\
  }\bibfield  {title} {\enquote {\bibinfo {title} {{Measurement of
  transverse-single-spin asymmetries for midrapidity and forward-rapidity
  production of hadrons in polarized p+p collisions at $\sqrt{s}=$200 and 62.4
  GeV}},}\ }\href {\doibase 10.1103/PhysRevD.90.012006} {\bibfield  {journal}
  {\bibinfo  {journal} {Phys. Rev. D}\ }\textbf {\bibinfo {volume} {90}},\
  \bibinfo {pages} {012006} (\bibinfo {year} {2014})}\BibitemShut {NoStop}%
\bibitem [{\citenamefont {Kopeliovich}\ \emph {et~al.}(2011)\citenamefont
  {Kopeliovich}, \citenamefont {Potashnikova}, \citenamefont {Schmidt},\ and\
  \citenamefont {Soffer}}]{Kopeliovich:2011bx}%
  \BibitemOpen
  \bibfield  {author} {\bibinfo {author} {\bibfnamefont {B.Z.}\ \bibnamefont
  {Kopeliovich}}, \bibinfo {author} {\bibfnamefont {I.K.}\ \bibnamefont
  {Potashnikova}}, \bibinfo {author} {\bibfnamefont {I.}~\bibnamefont
  {Schmidt}}, \ and\ \bibinfo {author} {\bibfnamefont {J.}~\bibnamefont
  {Soffer}},\ }\bibfield  {title} {\enquote {\bibinfo {title} {{Single
  transverse spin asymmetry of forward neutrons}},}\ }\href {\doibase
  10.1103/PhysRevD.84.114012} {\bibfield  {journal} {\bibinfo  {journal} {Phys.
  Rev. D}\ }\textbf {\bibinfo {volume} {84}},\ \bibinfo {pages} {114012}
  (\bibinfo {year} {2011})}\BibitemShut {NoStop}%
\bibitem [{\citenamefont {Aidala}\ \emph {et~al.}(2018)\citenamefont {Aidala}
  \emph {et~al.}}]{Aidala:2017cnz}%
  \BibitemOpen
  \bibfield  {author} {\bibinfo {author} {\bibfnamefont {C.}~\bibnamefont
  {Aidala}} \emph {et~al.} (\bibinfo {collaboration} {PHENIX Collaboration}),\
  }\bibfield  {title} {\enquote {\bibinfo {title} {{Nuclear Dependence of the
  Transverse-Single-Spin Asymmetry for Forward Neutron Production in Polarized
  $p+A$ Collisions at $\sqrt{{s}_{NN}}=200$ GeV}},}\ }\href {\doibase
  10.1103/PhysRevLett.120.022001} {\bibfield  {journal} {\bibinfo  {journal}
  {Phys. Rev. Lett.}\ }\textbf {\bibinfo {volume} {120}},\ \bibinfo {pages}
  {022001} (\bibinfo {year} {2018})}\BibitemShut {NoStop}%
\bibitem [{\citenamefont {Mitsuka}(2017)}]{Mitsuka:2017czj}%
  \BibitemOpen
  \bibfield  {author} {\bibinfo {author} {\bibfnamefont {G.}~\bibnamefont
  {Mitsuka}},\ }\bibfield  {title} {\enquote {\bibinfo {title} {{Recently
  measured large $A_N$ for forward neutrons in $p\uparrow{A}$ collisions at
  $\sqrt{s_{NN}}=200$ GeV explained through simulations of ultraperipheral
  collisions and hadronic interactions}},}\ }\href {\doibase
  10.1103/PhysRevC.95.044908} {\bibfield  {journal} {\bibinfo  {journal} {Phys.
  Rev. C}\ }\textbf {\bibinfo {volume} {95}},\ \bibinfo {pages} {044908}
  (\bibinfo {year} {2017})}\BibitemShut {NoStop}%
\bibitem [{\citenamefont {Acharya}\ \emph {et~al.}(2021)\citenamefont {Acharya}
  \emph {et~al.}}]{Acharya:2020opv}%
  \BibitemOpen
  \bibfield  {author} {\bibinfo {author} {\bibfnamefont {U.~A.}\ \bibnamefont
  {Acharya}} \emph {et~al.} (\bibinfo {collaboration} {PHENIX Collaboration}),\
  }\bibfield  {title} {\enquote {\bibinfo {title} {{Transverse momentum
  dependent forward neutron single spin asymmetries in transversely polarized
  $p+p$ collisions at $\sqrt {s}$ = 200 GeV}},}\ }\href {\doibase
  10.1103/PhysRevD.103.032007} {\bibfield  {journal} {\bibinfo  {journal}
  {Phys. Rev. D}\ }\textbf {\bibinfo {volume} {103}},\ \bibinfo {pages}
  {032007} (\bibinfo {year} {2021})}\BibitemShut {NoStop}%
\bibitem [{\citenamefont {Adcox}\ \emph {et~al.}(2003)\citenamefont {Adcox}
  \emph {et~al.}}]{Adcox:2003zm}%
  \BibitemOpen
  \bibfield  {author} {\bibinfo {author} {\bibfnamefont {K.}~\bibnamefont
  {Adcox}} \emph {et~al.} (\bibinfo {collaboration} {PHENIX Collaboration}),\
  }\bibfield  {title} {\enquote {\bibinfo {title} {{PHENIX detector
  overview}},}\ }\href {\doibase 10.1016/S0168-9002(02)01950-2} {\bibfield
  {journal} {\bibinfo  {journal} {Nucl. Instrum. Methods Phys. Res., Sec. A}\
  }\textbf {\bibinfo {volume} {499}},\ \bibinfo {pages} {469} (\bibinfo {year}
  {2003})}\BibitemShut {NoStop}%
\bibitem [{\citenamefont {Brun}\ \emph {et~al.}(1994)\citenamefont {Brun},
  \citenamefont {Bruyant}, \citenamefont {Carminati}, \citenamefont {Giani},
  \citenamefont {Maire}, \citenamefont {McPherson}, \citenamefont {Patrick},\
  and\ \citenamefont {Urban}}]{Brun:1994aa}%
  \BibitemOpen
  \bibfield  {author} {\bibinfo {author} {\bibfnamefont {R.}~\bibnamefont
  {Brun}}, \bibinfo {author} {\bibfnamefont {F.}~\bibnamefont {Bruyant}},
  \bibinfo {author} {\bibfnamefont {F.}~\bibnamefont {Carminati}}, \bibinfo
  {author} {\bibfnamefont {S.}~\bibnamefont {Giani}}, \bibinfo {author}
  {\bibfnamefont {M.}~\bibnamefont {Maire}}, \bibinfo {author} {\bibfnamefont
  {A.}~\bibnamefont {McPherson}}, \bibinfo {author} {\bibfnamefont
  {G.}~\bibnamefont {Patrick}}, \ and\ \bibinfo {author} {\bibfnamefont
  {L.}~\bibnamefont {Urban}},\ }\href {\doibase 10.17181/CERN.MUHF.DMJ1} {\emph
  {\bibinfo {title} {{GEANT Detector Description and Simulation Tool}}}}
  (\bibinfo {year} {1994})\BibitemShut {NoStop}%
\bibitem [{\citenamefont {Roesler}\ \emph {et~al.}(2000)\citenamefont
  {Roesler}, \citenamefont {Engel},\ and\ \citenamefont
  {Ranft}}]{Roesler:2000he}%
  \BibitemOpen
  \bibfield  {author} {\bibinfo {author} {\bibfnamefont {S.}~\bibnamefont
  {Roesler}}, \bibinfo {author} {\bibfnamefont {R.}~\bibnamefont {Engel}}, \
  and\ \bibinfo {author} {\bibfnamefont {J.}~\bibnamefont {Ranft}},\ }\bibfield
   {title} {\enquote {\bibinfo {title} {{The Monte Carlo event generator
  DPMJET-III}},}\ }in\ \href {\doibase 10.1007/978-3-642-18211-2_166} {\emph
  {\bibinfo {booktitle} {{International Conference on Advanced Monte Carlo for
  Radiation Physics, Particle Transport Simulation and Applications (MC
  2000)}}}}\ (\bibinfo {year} {2000})\ p.\ \bibinfo {pages} {1033},\ \Eprint
  {http://arxiv.org/abs/hep-ph/0012252} {arXiv:hep-ph/0012252} \BibitemShut
  {NoStop}%
\bibitem [{\citenamefont {Sj\"ostrand}\ \emph {et~al.}(2001)\citenamefont
  {Sj\"ostrand}, \citenamefont {L\"onnblad},\ and\ \citenamefont
  {Mrenna}}]{Sjostrand:2001yu}%
  \BibitemOpen
  \bibfield  {author} {\bibinfo {author} {\bibfnamefont {T.}~\bibnamefont
  {Sj\"ostrand}}, \bibinfo {author} {\bibfnamefont {L.}~\bibnamefont
  {L\"onnblad}}, \ and\ \bibinfo {author} {\bibfnamefont {S.}~\bibnamefont
  {Mrenna}},\ }\href@noop {} {\emph {\bibinfo {title} {{PYTHIA 6.2: Physics and
  manual}}}} (\bibinfo {year} {2001}),\ \Eprint
  {http://arxiv.org/abs/hep-ph/0108264} {arXiv:hep-ph/0108264} \BibitemShut
  {NoStop}%
\bibitem [{\citenamefont {Sj\"ostrand}\ \emph {et~al.}(2015)\citenamefont
  {Sj\"ostrand}, \citenamefont {Ask}, \citenamefont {Christiansen},
  \citenamefont {Corke}, \citenamefont {Desai}, \citenamefont {Ilten},
  \citenamefont {Mrenna}, \citenamefont {Prestel}, \citenamefont {Rasmussen},\
  and\ \citenamefont {Skands}}]{Sjostrand:2014zea}%
  \BibitemOpen
  \bibfield  {author} {\bibinfo {author} {\bibfnamefont {T.}~\bibnamefont
  {Sj\"ostrand}}, \bibinfo {author} {\bibfnamefont {S.}~\bibnamefont {Ask}},
  \bibinfo {author} {\bibfnamefont {J.~R.}\ \bibnamefont {Christiansen}},
  \bibinfo {author} {\bibfnamefont {R.}~\bibnamefont {Corke}}, \bibinfo
  {author} {\bibfnamefont {N.}~\bibnamefont {Desai}}, \bibinfo {author}
  {\bibfnamefont {P.}~\bibnamefont {Ilten}}, \bibinfo {author} {\bibfnamefont
  {S.}~\bibnamefont {Mrenna}}, \bibinfo {author} {\bibfnamefont
  {S.}~\bibnamefont {Prestel}}, \bibinfo {author} {\bibfnamefont {C.~O.}\
  \bibnamefont {Rasmussen}}, \ and\ \bibinfo {author} {\bibfnamefont {P.~Z.}\
  \bibnamefont {Skands}},\ }\bibfield  {title} {\enquote {\bibinfo {title} {{An
  introduction to PYTHIA 8.2}},}\ }\href {\doibase 10.1016/j.cpc.2015.01.024}
  {\bibfield  {journal} {\bibinfo  {journal} {Comput. Phys. Commun.}\ }\textbf
  {\bibinfo {volume} {191}},\ \bibinfo {pages} {159} (\bibinfo {year}
  {2015})}\BibitemShut {NoStop}%
\bibitem [{\citenamefont {Klein}\ \emph {et~al.}(2017)\citenamefont {Klein},
  \citenamefont {Nystrand}, \citenamefont {Seger}, \citenamefont {Gorbunov},\
  and\ \citenamefont {Butterworth}}]{Klein:2016yzr}%
  \BibitemOpen
  \bibfield  {author} {\bibinfo {author} {\bibfnamefont {S.~R.}\ \bibnamefont
  {Klein}}, \bibinfo {author} {\bibfnamefont {J.}~\bibnamefont {Nystrand}},
  \bibinfo {author} {\bibfnamefont {J.}~\bibnamefont {Seger}}, \bibinfo
  {author} {\bibfnamefont {Y.}~\bibnamefont {Gorbunov}}, \ and\ \bibinfo
  {author} {\bibfnamefont {J.}~\bibnamefont {Butterworth}},\ }\bibfield
  {title} {\enquote {\bibinfo {title} {{STARlight: A Monte Carlo simulation
  program for ultra-peripheral collisions of relativistic ions}},}\ }\href
  {\doibase 10.1016/j.cpc.2016.10.016} {\bibfield  {journal} {\bibinfo
  {journal} {Comput. Phys. Commun.}\ }\textbf {\bibinfo {volume} {212}},\
  \bibinfo {pages} {258} (\bibinfo {year} {2017})}\BibitemShut {NoStop}%
\bibitem [{\citenamefont {D'Agostini}(1995)}]{DAgostini:1994fjx}%
  \BibitemOpen
  \bibfield  {author} {\bibinfo {author} {\bibfnamefont {G.}~\bibnamefont
  {D'Agostini}},\ }\bibfield  {title} {\enquote {\bibinfo {title} {{A
  multidimensional unfolding method based on Bayes' theorem}},}\ }\href
  {\doibase 10.1016/0168-9002(95)00274-X} {\bibfield  {journal} {\bibinfo
  {journal} {Nucl. Instrum. Methods Phys. Res., Sec. A}\ }\textbf {\bibinfo
  {volume} {362}},\ \bibinfo {pages} {487} (\bibinfo {year}
  {1995})}\BibitemShut {NoStop}%
\bibitem [{\citenamefont {Adye}(2011)}]{Adye:2011gm}%
  \BibitemOpen
  \bibfield  {author} {\bibinfo {author} {\bibfnamefont {T.}~\bibnamefont
  {Adye}},\ }\bibfield  {title} {\enquote {\bibinfo {title} {{Unfolding
  algorithms and tests using RooUnfold}},}\ }in\ \href {\doibase
  10.5170/CERN-2011-006.313} {\emph {\bibinfo {booktitle} {{PHYSTAT 2011}}}}\
  (\bibinfo  {publisher} {CERN},\ \bibinfo {address} {Geneva},\ \bibinfo {year}
  {2011})\ p.\ \bibinfo {pages} {313},\ \Eprint
  {http://arxiv.org/abs/1105.1160} {arXiv:1105.1160 [physics.data-an]}
  \BibitemShut {NoStop}%
\bibitem [{\citenamefont {Brun}\ and\ \citenamefont
  {Rademakers}(1997)}]{Brun:1997pa}%
  \BibitemOpen
  \bibfield  {author} {\bibinfo {author} {\bibfnamefont {R.}~\bibnamefont
  {Brun}}\ and\ \bibinfo {author} {\bibfnamefont {F.}~\bibnamefont
  {Rademakers}},\ }\bibfield  {title} {\enquote {\bibinfo {title} {{ROOT: An
  object oriented data analysis framework}},}\ }\href {\doibase
  10.1016/S0168-9002(97)00048-X} {\bibfield  {journal} {\bibinfo  {journal}
  {Nucl. Instrum. Methods Phys. Res., Sec. A}\ }\textbf {\bibinfo {volume}
  {389}},\ \bibinfo {pages} {81} (\bibinfo {year} {1997})}\BibitemShut
  {NoStop}%
\bibitem [{\citenamefont {H{\"o}cker}\ and\ \citenamefont
  {Kartvelishvili}(1996)}]{Hocker:1995kb}%
  \BibitemOpen
  \bibfield  {author} {\bibinfo {author} {\bibfnamefont {A.}~\bibnamefont
  {H{\"o}cker}}\ and\ \bibinfo {author} {\bibfnamefont {V.}~\bibnamefont
  {Kartvelishvili}},\ }\bibfield  {title} {\enquote {\bibinfo {title} {{SVD
  approach to data unfolding}},}\ }\href {\doibase
  10.1016/0168-9002(95)01478-0} {\bibfield  {journal} {\bibinfo  {journal}
  {Nucl. Instrum. Methods Phys. Res., Sec. A}\ }\textbf {\bibinfo {volume}
  {372}},\ \bibinfo {pages} {469} (\bibinfo {year} {1996})}\BibitemShut
  {NoStop}%
\bibitem [{pol(2018)}]{polarimetry}%
  \BibitemOpen
  \href@noop {} {\enquote {\bibinfo {title} {{The RHIC polarimetry group}, note
  no 490},}\ } (\bibinfo {year} {2018})\BibitemShut {NoStop}%
\bibitem [{\citenamefont {Mitsuka}(2021)}]{gakufine}%
  \BibitemOpen
  \bibfield  {author} {\bibinfo {author} {\bibfnamefont {Gaku}\ \bibnamefont
  {Mitsuka}},\ }\href@noop {} {\enquote {\bibinfo {title} {Private
  communication based on~\cite{Mitsuka:2017czj}},}\ } (\bibinfo {year}
  {2021})\BibitemShut {NoStop}%
\bibitem [{\citenamefont {Guzey}\ and\ \citenamefont
  {Strikman}(2006)}]{Guzey:2005tk}%
  \BibitemOpen
  \bibfield  {author} {\bibinfo {author} {\bibfnamefont {V.}~\bibnamefont
  {Guzey}}\ and\ \bibinfo {author} {\bibfnamefont {M.}~\bibnamefont
  {Strikman}},\ }\bibfield  {title} {\enquote {\bibinfo {title}
  {{Proton-nucleus scattering and cross section fluctuations at RHIC and
  LHC}},}\ }\href {\doibase 10.1016/j.physletb.2005.11.065} {\bibfield
  {journal} {\bibinfo  {journal} {Phys. Lett. B}\ }\textbf {\bibinfo {volume}
  {633}},\ \bibinfo {pages} {245} (\bibinfo {year} {2006})}\BibitemShut
  {NoStop}%
\bibitem [{\citenamefont {Mitsuka}(2015)}]{Mitsuka:2015twa}%
  \BibitemOpen
  \bibfield  {author} {\bibinfo {author} {\bibfnamefont {G.}~\bibnamefont
  {Mitsuka}},\ }\bibfield  {title} {\enquote {\bibinfo {title} {{Forward hadron
  production in ultra-peripheral proton\textendash{}heavy-ion collisions at the
  LHC and RHIC}},}\ }\href {\doibase 10.1140/epjc/s10052-015-3848-0} {\bibfield
   {journal} {\bibinfo  {journal} {Eur. Phys. J. C}\ }\textbf {\bibinfo
  {volume} {75}},\ \bibinfo {pages} {614} (\bibinfo {year} {2015})}\BibitemShut
  {NoStop}%
\bibitem [{\citenamefont {Drechsel}\ \emph {et~al.}(2007)\citenamefont
  {Drechsel}, \citenamefont {Kamalov},\ and\ \citenamefont
  {Tiator}}]{Drechsel:2007if}%
  \BibitemOpen
  \bibfield  {author} {\bibinfo {author} {\bibfnamefont {D.}~\bibnamefont
  {Drechsel}}, \bibinfo {author} {\bibfnamefont {S.~S.}\ \bibnamefont
  {Kamalov}}, \ and\ \bibinfo {author} {\bibfnamefont {L.}~\bibnamefont
  {Tiator}},\ }\bibfield  {title} {\enquote {\bibinfo {title} {{Unitary Isobar
  Model- MAID2007}},}\ }\href {\doibase 10.1140/epja/i2007-10490-6} {\bibfield
   {journal} {\bibinfo  {journal} {Eur. Phys. J. A}\ }\textbf {\bibinfo
  {volume} {34}},\ \bibinfo {pages} {69} (\bibinfo {year} {2007})}\BibitemShut
  {NoStop}%
\end{thebibliography}

%
 
\end{document}